\documentclass[12pt,preprint]{aastex}
\begin{document}
\title{Fast and Accurate Fourier Series Solutions to Gravitational Lensing 
by A General Family of Two Power-Law Mass Distributions}

\author{Kyu-Hyun Chae}

\affil{University of Manchester, 
Jodrell Bank Observatory, Macclesfield, Cheshire
SK11 9DL, U.K.; University of Pittsburgh, 
Department of Physics and Astronomy, Pittsburgh, PA 15260, U.S.A.}

\begin{abstract}
Fourier series solutions to the deflection and magnification by a family of 
three-dimensional cusped two power-law ellipsoidal mass distributions are 
presented. The cusped two power-law ellipsoidal mass distributions are 
characterized by inner and outer power-law radial indices and a break 
(or, transition) radius. The model family
includes mass models mimicking Jaffe, Hernquist, and $\eta$ models and dark 
matter halo profiles from numerical simulations. The Fourier series solutions
for the cusped two power-law mass distributions are relatively simple, 
and allow a very fast calculation even for a chosen small fractional
calculational error (e.g.\ $10^{-5}$). These results will be particularly 
useful for studying lensed systems which provide a number of accurate lensing
constraints and for systematic analyses of large numbers of lenses. 
Subroutines employing these results for the two power-law model and the 
results by Chae, Khersonsky, \& Turnshek for the generalized single 
power-law mass model are made publicly available.
\end{abstract}


\section{Introduction}
Strong gravitational lensing effects provide us with direct tools to probe 
the potentials of lensing objects (i.e.\ galaxies and clusters/groups of
galaxies) [see, e.g., Schneider, Ehlers, \& Falco (1992) for an introduction].
Strong gravitational lensing effects can also be used to constrain cosmological
models, in particular to directly measure the Hubble constant ($H_0$) 
in a standard cosmology via light travel time differences between 
 lensed images of an extragalactic variable source (Refsdal 1964).
Lensed image properties are related to (and hence probe) the angular 
structures and radial behaviors of lensing mass distributions.\footnote{
Gravitational lensing probes the total mass distribution of a lensing object, 
namely the sum of the luminous and dark mass distributions.}
The angular structures and radial behaviors of lensing objects are 
usually studied using suitable parametric mass models. The choice of a mass
model in lensing studies depends on two practical issues, apart from the 
question of whether the candidate model can be a good approximation to the
``true'' lensing mass distributions: 
(1) whether the parameters of the candidate model are constrainable
with the available lensing data; (2) the availability of a method
which can be used to calculate the deflection and magnification of 
the candidate model with the speed and accuracy required for the problems
under consideration.

This paper is concerned with the second issue for 
a family of three-dimensional triaxial cusped two 
power-law mass distributions, which are described by $\rho(R) =
\rho_0 R^{-\nu_i} (1 + R^2)^{-(\nu_o-\nu_i)/2}$ ($R^2 = x^2/a^2 + y^2/b^2
+ z^2/c^2$) (Mu\~{n}oz, Kochanek, \& Keeton 2001). This family of mass 
distributions include mass models resembling well-known galaxy models 
(e.g.\ Jaffe 1983, Hernquist 1990, Dehnen 1993, Tremaine et al. 1994, Zhao 
1996) and theoretical dark matter halo profiles from numerical simulations 
(e.g.\ Navarro, Frenk, \& White 1997, Moore et al.\ 1998). 
These distributions allow more detailed studies of 
radial behaviors of galactic mass distributions via a break radius
and an outer power-law radial index (i.e.\ $\nu_o$ in the above), compared to
the generalized single power-law mass distributions (Chae, Khersonsky, \& 
Turnshek 1998 [hereafter CKT]; Barkana 1998). It is also more realistic than
the simpler two-dimensional double power-law models which were used by 
Evans \& Wilkinson (1998) to study analytical properties of cusped lens models.
To date, the only available method for calculating 
the deflection and magnification for this family of three-dimensional 
mass distributions has been based on numerically
evaluating a set of one-dimensional integrals
(Mu\~{n}oz et al.\ 2001; Keeton 2001). 
However, the numerical integration method is particularly slow for the
cusped two power-law model family because the surface density for the model
is not represented by a simple function (see \S 2).
This paper presents mathematical solutions to the 
deflection and magnification by the cusped two power-law model family.
The obtained mathematical solutions allow accurate calculations with errors of
$10^{-6}$ or smaller for most of the parameter space and they improve 
the calculation speed dramatically for a large portion of the parameter space.

This paper is organized as follows. In \S 2, we review the mathematical 
expressions for the projected surface mass densities of the three-dimensional
cusped two power-law mass model and the two other models which are necessary
to build up the model. 
In \S 3, the Fourier series formalism of CKT is reviewed and Fourier series
coefficient functions of the three-dimensional cusped two power-law mass model
are obtained for two parameter subspaces for which the radius is,
respectively, smaller and larger than a break radius scaled with an 
ellipticity; mathematical details are outlined in the Appendices. 
A discussion of the results is given in \S 4.

\section{The Projected Surface Mass Density for the Cusped Two Power-law Model}
The three-dimensional cusped two power-law ellipsoidal mass model, referred to
as `cusp' model, is given by
\begin{equation}
\rho^{\mbox{\scriptsize cusp}}(X,Y,Z) = \rho_0 R^{-\nu_i} 
(1 + R^2)^{-(\nu_o-\nu_i)/2} \hspace{1em} \left[ R^2 = 
\left(\frac{X}{a}\right)^2 +  \left(\frac{Y}{b}\right)^2 + 
\left(\frac{Z}{c}\right)^2 
 \right]
\end{equation}
where $\nu_i$ and $\nu_o$ are inner and outer power-law radial indices 
respectively, and ($X$, $Y$, $Z$) and ($a$, $b$, $c$) are body coordinates 
and ``break sizes'' along the axes, respectively. 

The surface mass density for the `cusp' mass model projected on 
the lens plane is, up to a multiplicative constant controlling the deflection
scale of the lens,\footnote{The multiplicative constant, denoted by $\kappa_0$,
is related to the critical radius of the lens for $e = 0$, denoted by $r_E$,
by the following equation, 
\[ \kappa_0 = \frac{1}{2} \xi^2 \left[{\mbox{B}}\left(\frac{\nu_o-3}{2},
 \frac{3-\nu_i}{2}\right)\right]^{-1} J^{-1}, \]
with
\begin{eqnarray*}
J & = & 1 - \left( \frac{1}{1 + \xi^2} \right)^{(\nu_o-3)/2} 
\left[ F\left(\frac{\nu_i}{2},\frac{\nu_o-3}{2};\frac{\nu_i-1}{2}; 
\frac{\xi^2}{1+\xi^2} \right) \right. \\
 &  & + \left( \frac{\xi^2}{1+\xi^2}\right)^{(3-\nu_i)/2} 
\frac{\Gamma(3/2) \Gamma[(\nu_o-\nu_i)/2] \Gamma[(\nu_i-3)/2]}
 {\Gamma(\nu_i/2) \Gamma[(3-\nu_i)/2]  \Gamma[(\nu_o-3)/2]} \\
 &  & \left. \times F\left(\frac{3}{2},\frac{\nu_o-\nu_i}{2};\frac{5-\nu_i}{2};
 \frac{\xi^2}{1+\xi^2} \right) \right],
\end{eqnarray*}
where $\xi = r_E/r_b$. }
 given by
\begin{mathletters}
\begin{equation}
\kappa^{\mbox{\scriptsize cusp}}(\zeta) = 
  {\mbox{B}}\left(\frac{1}{2},\frac{\nu_o-1}{2}\right)
    \frac{1}{(1+\zeta^2)^{(\nu_o-1)/2}}
  F\left(\frac{\nu_i}{2},\frac{\nu_o-1}{2};\frac{\nu_o}{2};\frac{1}{1+\zeta^2}
   \right),
\end{equation}
which is, for $\nu_i \neq 1$,  equivalent to
\begin{eqnarray}
\kappa^{\mbox{\scriptsize cusp}}(\zeta) &= &
  {\mbox{B}}\left(\frac{1}{2},\frac{\nu_i-1}{2}\right)
     \frac{\zeta^{-(\nu_i -1)}}{(1 + \zeta^2)^{(\nu_o-\nu_i)/2}}
  F\left(\frac{\nu_o-\nu_i}{2},\frac{1}{2};\frac{3-\nu_i}{2};
    \frac{\zeta^2}{1+\zeta^2} \right) + \nonumber \\
    &  & {\mbox{B}}\left(\frac{\nu_o-1}{2},\frac{1-\nu_i}{2}\right)
     \frac{1}{(1 + \zeta^2)^{(\nu_o-1)/2}}
  F\left(\frac{\nu_i}{2},\frac{\nu_o-1}{2};\frac{\nu_i+1}{2};
     \frac{\zeta^2}{1+\zeta^2} \right),
\end{eqnarray}
\end{mathletters}
where
\begin{equation}
\zeta = \frac{r}{r_b}[1 + e \cos 2(\phi - \phi_0)]^{1/2},
\end{equation}
where parameter $e (\ge 0)$ is related to the ellipticity ($\epsilon$)
via $\epsilon = 1 - [(1-e)/(1+e)]^{1/2}$. In equations (2a) and (2b), 
B($a,b$) is the beta function and $F(a,b;c;z)$ is the hypergeometric function.
The relations between the parameters of the surface mass density [eqs.\ (2a) 
\& (2b)] and the parameters of the three-dimensional mass density 
[eq.\ (1)] and three Eulerian angles can be obtained, e.g., following CKT.
The first expression given by equation (2a) is suitable for calculating
the deflection and magnification for $r \ge r_b/(1+e)^{1/2}$, while
the second expression given by equation (2b) is suitable for
$r < r_b/(1+e)^{1/2}$. 

Each term in (the expansion of) the right-hand side of equation (2a) is 
the surface mass density of the familiar softened
general single power-law mass model (CKT; Barkana 1998) given by
 \begin{equation}
\kappa^{\mbox{\scriptsize sple}}(\zeta;\mu) =
 \frac{1}{(1+\zeta^2)^{\mu+1}},
\end{equation}
for which Fourier series solutions for the
deflection and magnification are given in CKT. As indicated by a superscript,
the surface density of equation (4) is referred to as `sple' (i.e.\ softened
power-law ellipsoid) model.
Equation (2b) consists of terms of the following form, 
which will be referred to as `Nuker' model following the {\it Hubble Space
Telescope} group who observed the centers of nearby early-type galaxies (e.g.\
Lauer et al.\ 1995), although the inner power-law index 
[$2\lambda$ in eq.\ (5)] can take positive values here, 
\begin{equation}
\kappa^{\mbox{\scriptsize Nuker}}(\zeta;\lambda,\mu) = 
\frac{\zeta^{2\lambda}}{(1 +\zeta^2)^{\mu + 1}},
\end{equation}
where $\lambda > -1$.  Note that multiplicative constants were suppressed
in both equations (4) and (5) for mathematical convenience, as in
equations (2a) and (2b).  For positive integer values
of $\lambda$, equation (5) can be written as
\begin{equation}
\kappa^{\mbox{\scriptsize Nuker}}(\zeta;\lambda,\mu) = 
\sum_{l=0}^{\lambda} (-1)^l \left( \begin{array}{c} \lambda \\ l \end{array}
  \right) \kappa^{\mbox{\scriptsize sple}}(\zeta;\mu-\lambda+l).
\end{equation}
Summary of the mass models given above can be found in
Table~1.

\section{Fourier Series Coefficients for `CUSP' Model}
As was noticed in CKT for the `sple' model [eq.\ (4)], the surface potential,
deflection and magnification for a surface mass density with elliptical 
symmetry [i.e., which is a function of $\zeta$ given by equation (3)] can be 
written as Fourier series with coefficients which can be written as linear 
combinations of the following ``$I$-functions''\footnote{These $I$-functions
are constants for singular isothermal mass distributions.} of $r$ whose 
general definitions are given in Appendix A: $I^{(0)}(r)$, 
$I^{(0')}(r)$,\footnote{This symbol is used throughout in place of 
$I'^{(0)}(r)$ used in CKT.}  
$I_{2m}^{(1)}(r)$, $I_{2m}^{(2)}(r)$, and $I_{2m}^{(3)}(r)$  
($m=1,2,3,\ldots$).\footnote{For $m=0$, $I^{(0)}(r) = I_{2m}^{(1)}(r)$
and $I^{(0')}(r) = 2 I_{2m}^{(3)}(r)$.} 
 Using these $I$-functions, we may write the surface 
potential, $\psi = \psi(r,\phi)$,  and its first
and second derivatives  in polar coordinates as follows:
\begin{eqnarray}
\psi & = & \int^r I^{(0)}(r') dr'  -
\sum_{m=1}^{\infty} \frac{r}{2m}[I_{2m}^{(1)}(r) + I_{2m}^{(2)}(r)]
 \cos [2m(\phi - \phi_0)],  \\
\frac{\partial \psi}{\partial r} & = & 
 I^{(0)}(r) + \sum_{m=1}^{\infty} [I_{2m}^{(1)}(r) - I_{2m}^{(2)}(r)]
 \cos [2m(\phi - \phi_0)],  \\
\frac{\partial \psi}{r \partial\phi} & = & 
 \sum_{m=1}^{\infty} [I_{2m}^{(1)}(r) + I_{2m}^{(2)}(r)] 
 \sin [2m(\phi - \phi_0)],  \\
\frac{\partial}{\partial r}\left(\frac{\partial \psi}{\partial r}
\right) & = & 
\frac{-I^{(0)}(r)+I^{(0')}(r)}{r} \nonumber \\
 &  & - \frac{1}{r} \sum_{m=1}^{\infty}
 [(2m+1)I_{2m}^{(1)}(r)+(2m-1)I_{2m}^{(2)}(r)-4 I_{2m}^{(3)}(r)]
 \cos [2m(\phi - \phi_0)], \nonumber \\
  &  &    \\
\frac{\partial}{r \partial \phi}\left(\frac{\partial \psi}{\partial r}
\right) & = &  
-\frac{2}{r} \sum_{m=1}^{\infty} m [I_{2m}^{(1)}(r) - I_{2m}^{(2)}(r)]
 \sin [2m(\phi - \phi_0)], \\
\frac{\partial}{\partial r}\left(\frac{\partial \psi}{r \partial \phi}
\right) & = & -\frac{1}{r} \sum_{m=1}^{\infty} 
[(2m+1)I_{2m}^{(1)}(r)-(2m-1)I_{2m}^{(2)}(r)] \sin [2m(\phi - \phi_0)],  \\
\frac{\partial}{r \partial \phi}
\left(\frac{\partial \psi}{r \partial \phi} \right) & = &
\frac{2}{r} \sum_{m=1}^{\infty} m [I_{2m}^{(1)}(r) + I_{2m}^{(2)}(r)]
 \cos [2m(\phi - \phi_0)].
\end{eqnarray}
As shown above, the $m$-th order Fourier coefficients of the potential, its
first and second derivatives are entirely determined by the $m$-th order 
$I$-functions. For any surface mass density with elliptical symmetry and
a reasonable ellipticity, the magnitude of an $I$-function decreases rapidly
as order $m$ increases. The $I^{(1)}(r)$ and $I^{(2)}(r)$ functions are
determined solely by the mass distributions within and outside the radius $r$,
respectively. The $I^{(3)}(r)$ functions arise in the derivatives of the 
$I^{(1)}(r)$ and $I^{(2)}(r)$ functions.  

Below we obtain mathematical expressions for the functions $I^{(0)}(r)$, 
$I^{(0')}(r)$, $I_{2m}^{(1)}(r)$, $I_{2m}^{(2)}(r)$, and $I_{2m}^{(3)}(r)$ 
of the `cusp' model using the two expressions of the surface density 
[equations (2a) \& (2b)]. As was mentioned in \S 2, the surface density 
expressions of the `cusp' model consist of those of the `sple' and `Nuker'
models. Thus, in order to calculate the $I$-functions of `cusp' model,
we need the $I$-functions of the `sple' and `Nuker' models.
Mathematical details outlining the derivations of the necessary $I$-functions 
of the `sple' and `Nuker' models are given in the Appendices A and B.

\subsection{The Case $r < r_b/(1+e)^{1/2}$}
From equations (2b) and (5), we have
\begin{eqnarray}
I^{\mbox{\scriptsize cusp}(i)}_{2m}(r) & = &\sum_{n=0}^{\infty} 
 [B_n^{(1)} (\nu_i,\nu_o) 
I^{\mbox{\scriptsize Nuker}(i)}_{2m}(r; -(\nu_i-1)/2 +n, (\nu_o-\nu_i)/2+n-1)
 \nonumber \\ &  & \hspace{2em} + B_n^{(2)}(\nu_i,\nu_o) 
I^{\mbox{\scriptsize Nuker}(i)}_{2m}(r; n, (\nu_o-3)/2+n)], \nonumber \\
  &   &   (i=1,2,3) 
\end{eqnarray}
where
\begin{eqnarray}
 B_n^{(1)} (\nu_i,\nu_o) & = & {\mbox{B}}
 \left(\frac{1}{2},\frac{\nu_i-1}{2}\right)
  \frac{\left(\frac{1}{2}\right)_n \left(\frac{\nu_o-\nu_i}{2}\right)_n}
   {\left(\frac{3-\nu_i}{2}\right)_n n!},    \\ 
 B_n^{(2)} (\nu_i,\nu_o) & = & {\mbox{B}}
 \left(\frac{\nu_o-1}{2},\frac{1-\nu_i}{2}\right)
  \frac{\left(\frac{\nu_i}{2}\right)_n \left(\frac{\nu_o-1}{2}\right)_n}
   {\left(\frac{\nu_i+1}{2}\right)_n n!}, 
\end{eqnarray}
and $I^{\mbox{\scriptsize Nuker}(i)}_{2m}(r; \lambda, \mu)$ are the 
$I$-functions of the `Nuker' model [eq.\ (5)]. In equations (15) and (16)
$(a)_b$ is the pochhammer symbol, i.e., $(a)_b = \Gamma(a+b)/\Gamma(a)$.
The $I$-functions of the `Nuker' model are given below.

\noindent
{\bf a.} {\it $\lambda$ = non-integer}

As outlined in Appendix A, the $I$-functions of the `Nuker' model for 
non-integer values of $\lambda$ are given by the following simple expressions,
\begin{eqnarray}
I^{\mbox{\scriptsize Nuker}(i)}_{2m}(r;\lambda,\mu) & = & r [-\sqrt{e_1}]^m 
 \left[\left(\frac{r}{r_b}\right)^2 e_2 \right]^{\lambda} \nonumber \\
 &  & \times \sum_{k=0}^{\infty} C^{(i)}_{k}(m;\lambda,\mu) 
 \left[ \left(\frac{r}{r_b}\right)^2 e_2 \right]^k
  F(m-k-\lambda,-k-\lambda;m+1;e_1), \nonumber \\
 &  &  (i=1,2,3) 
\end{eqnarray}
where
\begin{eqnarray}
 e_1 & = & \frac{1 - \sqrt{1 - e^2}}{1 + \sqrt{1 - e^2}},  \\
 e_2 & = & \frac{1}{2} ( 1 + \sqrt{1 - e^2}),
\end{eqnarray}
\begin{eqnarray}
 C^{(1)}_k(m;\lambda,\mu) & = & \frac{(-1)^k (\mu+1)_k (\lambda+m+1)_k
 (-k-\lambda)_m}{(m+\lambda+1)(m+\lambda+2)_k k! m!}, \\
 C^{(2)}_k(m;\lambda,\mu) & = & \frac{(-1)^k (\mu+1)_k (\lambda-m+1)_k
 (-k-\lambda)_m}{(m-\lambda-1)(-m+\lambda+2)_k k! m!}, \\
 C^{(3)}_k(m;\lambda,\mu) & = & \frac{(-1)^k (\mu+1)_k 
 (-k-\lambda)_m}{k! m!}. 
\end{eqnarray}

\noindent
{\bf b.} {\it $\lambda$ = zero or positive integer}

For zero or positive integer values of $\lambda$, the $I$-functions can be
calculated using equation (6) and the $I$-functions of the sple model, i.e.
\begin{equation}
 I_{2m}^{\mbox{\scriptsize Nuker}(i)}(r;\lambda,\mu) = 
 \sum_{l=0}^{\lambda} (-1)^l \left(\begin{array}{c}
  \lambda \\ l\end{array}\right)
 I_{2m}^{\mbox{\scriptsize sple} (i)}(r;\mu-\lambda+l)
   \hspace{1em} (i=1,2,3),
\end{equation}
where $I_{2m}^{\mbox{\scriptsize sple} (i)}(r;\mu)$ are the $I$-functions for
the `sple' model given by equation (4).

\subsection{The Case $r \ge r_b/(1+e)^{1/2}$}
For this case, the $I$-functions are calculated using equation (2a), which may
be re-written as
\begin{equation}
\kappa^{\mbox{\scriptsize cusp}}(\zeta) =  \sum_{n=0}^{\infty} 
D_n(\nu_i,\nu_o) \kappa^{\mbox{\scriptsize sple}}(\zeta;n+\mu),
\end{equation}
where $\mu = (\nu_o -3)/2$, $\kappa^{\mbox{\scriptsize sple}}(\zeta;\mu)$
is given by equation (4), and
\begin{equation}
D_n(\nu_i,\nu_o) = {\mbox{B}}\left(\frac{1}{2},\frac{\nu_o-1}{2}\right)
  \frac{\left(\frac{\nu_i}{2}\right)_n \left(\frac{\nu_o-1}{2}\right)_n}
   {\left(\frac{\nu_o}{2}\right)_n n!}.
\end{equation}
Thus, the $I$-functions of the cusp model may be written as: 
\begin{equation}
I_{2m}^{{\mbox{\scriptsize cusp}}(i)}(r) = 
\sum_{n=0}^{\infty} D_n(\nu_i,\nu_o)
   I_{2m}^{{\mbox{\scriptsize sple}}(i)}(r;n+\mu) \hspace{1em} (i=1,2,3),
\end{equation}
where $I_{2m}^{{\mbox{\scriptsize sple}}(i)}(r;\mu)$ are the $I$-functions
of the `sple' model [eq.\ (4)] for $r \ge r_b/(1+e)^{1/2}$ which are
given in CKT. The series sum of equation (26) can be efficiently calculated
by substituting the series expressions for 
$I_{2m}^{{\mbox{\scriptsize sple}}(i)}(r;\mu)$ given in CKT into the equation,
and turning double infinite sums into  single infinite sums using the 
following rule,
\begin{equation}
\sum_{i=0}^{\infty} \sum_{k=0}^{\infty} (\ldots) = 
\sum_{j=0}^{\infty} \sum_{k=0}^{j} (\ldots),
\end{equation}
where $j = i+k$.

The following equations (28), (29), (30b), (31), and (32) are obtained from
equations (26b), (43), (27b), (28), and (44) of CKT, respectively using the
rule given by equation (27), while equation (30a) is obtained from a new 
expression for $I_{2m}^{{\mbox{\scriptsize sple}}(1)}(r;\mu)$ which is given
in Appendix B [eq.\ (B2)]:
\begin{eqnarray}
I^{\mbox{\scriptsize cusp}(0)}(r) & = & 
 \frac{r_b^2}{r} \frac{1}{\sqrt{1-e^2}} \sum_{k=0}^{\infty}
  \frac{D_k(\nu_i,\nu_o)}{k+\mu}  \nonumber \\
  & & -h(r) [\varepsilon_2(r)]^{\mu} \sum_{j=0}^{\infty} [\varepsilon_2(r)]^j
   F(-j-\mu,-j-\mu;1;\varepsilon_1(r)) 
  \sum_{k=0}^{j} \frac{D_k(\nu_i,\nu_o)}{k+\mu},
\end{eqnarray}
\begin{eqnarray}
I^{\mbox{\scriptsize cusp}(0')}(r) & = &
 2 h(r) [\varepsilon_2(r)]^{\mu} \sum_{j=0}^{\infty}
  D_j(\nu_i,\nu_o) [\varepsilon_2(r)]^j F(-j-\mu,-j-\mu;1;\varepsilon_1(r)),
\end{eqnarray}
\begin{mathletters}
\begin{eqnarray}
I_{2m}^{\mbox{\scriptsize cusp}(1)}(r) & = &
 \frac{r}{\pi} (-te)^m (1-t)^{\mu+1} \sum_{k=0}^{\infty}
 \frac{[1 + (-1)^k](te)^k }{\Gamma(k+m+1) (k+2m+1)} {\mbox{Icos}}(k+m,m)
 \nonumber \\
 &  & \times \sum_{l=0}^{\infty} D_l(\nu_i,\nu_o) (1-t)^l
    \frac{\Gamma(l+k+m+\mu+1)}{\Gamma(l+\mu+1)}
     \nonumber \\
 &  & \times  F(l+k+m+\mu+1,1;k+2m+2;t) \nonumber \\
 &  & \left( t \equiv \frac{(\frac{r}{r_b})^2}{1 + (\frac{r}{r_b})^2}; 
 \hspace{1em}  {\mbox{Icos}}(k,m) \equiv \int_0^{\pi/2} dx \cos^k x \cos mx
   \right)  \nonumber \\
 &   & [{\mbox{for}} \hspace{1em} r < r_b/(1 - e)^{1/2}], 
\end{eqnarray}
\begin{eqnarray}
I_{2m}^{\mbox{\scriptsize cusp}(1)}(r)&= &
 -h(r)[-\sqrt{\varepsilon_1(r)}]^m [\varepsilon_2(r)]^{\mu}
 \frac{1}{\Gamma(m+1)} \sum_{j=0}^{\infty} [\varepsilon_2(r)]^j 
\frac{\Gamma(m+j+\mu+1)}{\Gamma(-m+j+\mu+1)} \nonumber \\
 &  & \times F(m-j-\mu,-j-\mu;m+1;\varepsilon_1(r)) 
 \sum_{k=0}^{j} D_k(\nu_i,\nu_o) \frac{\Gamma(k+\mu-m)}{\Gamma(k+\mu+1)}
  \nonumber \\
 &  & + h(r)[-\sqrt{\varepsilon_1(r)} \varepsilon_2(r)]^m \frac{1}{\Gamma(m+1)}
     \left[ \sum_{k=0}^{\infty} D_k(\nu_i,\nu_o) \frac{\Gamma(k+\mu-m)}
      {\Gamma(k+\mu+1)} \right] \nonumber \\ 
 &  & \times \left[ \sum_{j=0}^{\infty} [\varepsilon_2(r)]^j
   \frac{\Gamma(j+2m+1)}{\Gamma(j+1)} F(-j,-j-m;m+1;\varepsilon_1(r)) \right]
   \nonumber \\
 &  & [{\mbox{for}} \hspace{1em} r \ge r_b/(1 - e)^{1/2}],
\end{eqnarray}
\end{mathletters}
\begin{eqnarray}
I_{2m}^{\mbox{\scriptsize cusp}(2)}(r) & = &
 h(r)[-\sqrt{\varepsilon_1(r)}]^m [\varepsilon_2(r)]^{\mu}
\frac{1}{\Gamma(m+1)} \nonumber \\
 &  & \times \sum_{j=0}^{\infty} [\varepsilon_2(r)]^j 
F(m-j-\mu,-j-\mu;m+1;\varepsilon_1(r))  \nonumber \\
 &  & \times \sum_{k=0}^{j} D_k(\nu_i,\nu_o)
\frac{\Gamma(k+m+\mu)}{\Gamma(k+\mu+1)},
\end{eqnarray}
\begin{eqnarray}
I_{2m}^{\mbox{\scriptsize cusp}(3)}(r) & = & 
 h(r)[-\sqrt{\varepsilon_1(r)}]^m 
[\varepsilon_2(r)]^{\mu} \frac{1}{\Gamma(m+1)} \nonumber \\
 &  &\times \sum_{j=0}^{\infty} [\varepsilon_2(r)]^j 
F(m-j-\mu,-j-\mu;m+1;\varepsilon_1(r)) \nonumber \\
 &  &\times D_j(\nu_i,\nu_o) \frac{\Gamma(j+m+\mu+1)}{\Gamma(j+\mu+1)}.
\end{eqnarray}
In equations (28), (29), (30b), (31), and (32), the functions $h(r)$, 
$\varepsilon_1(r)$, and $\varepsilon_2(r)$ are given as follows (CKT):
\begin{eqnarray}
h(r) & = & \frac{r}{\sqrt{(1+s^2)^2-(e s^2)^2}}, \\
\varepsilon_1(r) & = & \left[ 1 -\sqrt{1 -
\left(\frac{e s^2}{1 + s^2}\right)^2} \right] 
\left[ 1 +\sqrt{1 - \left(\frac{e s^2}
{1 + s^2}\right)^2} \right]^{-1}, \\
\varepsilon_2(r) & = & \frac{1}{2} \left[ 
\frac{1 + s^2}{(1+s^2)^2 - (e s^2)^2}
+ \frac{1}{\sqrt{(1+ s^2)^2 - (e s^2)^2}} \right] \\
 &  & \left( s \equiv \frac{r}{r_b}  \right). \nonumber
\end{eqnarray}
In equations (28) \& (30b), the constant infinite sum can be evaluated using
$F(a,b;c;1) = \Gamma(c) \Gamma(c-a-b) /[\Gamma(c-a) \Gamma(c-b)]$ as follows,
\begin{equation}
\sum_{k=0}^{\infty} D_k(\nu_i,\nu_o) \frac{\Gamma(k+\mu-m)}
      {\Gamma(k+\mu+1)} = \frac{\Gamma(1/2) \Gamma[(\nu_o-3-2m)/2] 
  \Gamma[(3+2m-\nu_i)/2] }{\Gamma[(\nu_o-\nu_i)/2] \Gamma[(3+2m)/2]},
\end{equation}
where $m$ is zero or positive integer.
Equation (30a) converges in a non-trivial fashion
and requires a careful treatment. The convergence property of equation (30a)
and a strategy for its efficient numerical evaluation are 
discussed in Appendix C.

\section{Discussion}
We have obtained Fourier series solutions to the deflection and magnification
by the three-dimensional cusped two power-law ellipsoidal mass (`cusp')
model [eq.\ (1) or eqs.\ (2a,b); see also Mu\~{n}oz et al.\ (2001)] by 
calculating the $I$-functions of the model whose general definitions are given
in Appendix A. The $I$-functions of the `cusp' model given in \S 3 are 
well-defined rapidly converging series and can be evaluated efficiently using 
recursion relations for gamma functions and those for hypergeometric functions
for the successive terms in the series (Abramowitz \& Stegun 1964;
Gradshteyn \& Ryzhik 1994). Using these $I$-functions, 
the deflection and magnification can be calculated 
quickly and accurately for arbitrary parameter values. In particular, for
 any value of the radius sufficiently smaller or larger than the break radius
the calculation is very fast, even for highly accurate calculation, e.g., with
fractional errors of $10^{-5}$ or smaller. For example, compared with that for
the `sple' model for a given value of the normalized radius $r/r_b$, the 
calculational speed for the `cusp' model is slower by factors of $\sim 3$
and $\sim 8$ respectively for $r/r_b = 10$ and 
0.1.\footnote{The calculational speed in the Fourier series approach 
(for any model) depends on the mass ellipticity for any value of $r/r_b$. 
Execution time increases approximately linearly with ellipticity for
a fixed series truncation criterion.}  It typically takes $\sim 10^{-5}$ to
$\sim 10^{-3}$ seconds on an AMD Athlon 1.1 GHz CPU and an UltraSparc II 400
MHz CPU to calculate the deflection and magnification with fractional 
calculational errors of $10^{-5}$ or smaller for any ellipticity 
$\epsilon \lesssim 0.7$. 

Based on the present results, it is only around the break radius (specifically
for $0.7 r_b/\sqrt{1+e} \lesssim r \lesssim 1.3r_b/\sqrt{1-e}$) that the 
calculation is limited, mainly due to difficulty in efficiently
evaluating an $I^{(1)}$-function given by equation (30a). 
There are two limitations:
First, for ellipticities larger than $\epsilon \sim 0.7$,
smallest possible fractional calculational errors are $\sim 10^{-5}$. 
Second, the calculation slows down significantly compared with calculations
outside the above range of radius; on the same machines and with the same 
limits on the fractional calculational error and ellipticity given above
execution time is typically $\sim 0.5 \times 10^{-3}$ to $\sim 10^{-2}$ 
seconds. These limitations, although not very significant, can be reduced 
if a new $I^{(1)}$-function simpler than equation (30a) is obtained in the 
future.

To summarize, the present Fourier series solutions to the deflection and 
magnification by the `cusp' model improve the calculational speed by factors
of tens for a large portion of the parameter space of the model, compared to
numerical integrations. For radii close to the break radius, the Fourier 
method still appears to be a few times faster than the numerical integration
method for relatively small ellipticities (e.g.\ $\epsilon \lesssim 0.5$). 
The present Fourier series solutions for the 'cusp' model can also 
be used to calculate very accurate the deflection and 
magnification with fractional calculational errors of $10^{-6}$ or smaller, 
for the entire parameter space but the small portion of the parameter
space specified above. Therefore, for practical applications of the `cusp'
model in which one is mainly interested in calculating the deflection and 
magnification at radii not too close to the break radius, the present Fourier
series solutions greatly reduce his/her numerical efforts.

Numerical implementation of the $I$-functions of the `cusp' model given in
this paper is relatively straightforward except for two of them which
are discussed in Appendix C. However, for the convenience of the lensing
community well-tested codes both for the `cusp' [eqs.\ (2a,b)] and `sple'
[eq.\ (4)] models are made publicly available on the following web page 
{\bf{\mbox http://www.jb.man.ac.uk/$\sim$chae}}. The codes for the `cusp'
and `sple' models are given the names of `cuspFS' and 
`spleFS' respectively, where `FS' stands for Fourier series.

The remarkable calculational speed for a chosen small calculational error and 
the easy control over the calculational error (or, speed), which the Fourier 
series solutions both to the `sple'  and `cusp' models demonstrate, 
suggest that the Fourier series method is the most efficient method for
calculating the deflection and magnification for a general class of mass
distributions. In particular, for any surface mass density which is a function
of $\zeta$ [eq.\ (3)], one only needs to calculate the $I$-functions of the 
model [eqs.\ (A1) through (A5)] to calculate both the deflection and 
magnification. As shown in CKT and in this paper, the $I$-functions both of 
the `sple' and `cusp' models are well-defined, relatively simple converging 
series. It would be expected  that for other classes 
of mass distributions which have not yet been explored,
similar relatively simple expressions for the $I$-functions exist. 
For example, a generalized softened power-law ellipsoid with an exponentially
declining envelope given by
\begin{equation}
\rho(R) = \frac{\rho_0}{(1 + R^2)^{\nu/2}} \exp(-R^2/R_t^2),
\end{equation}
where $R^2 = (X/a)^2 + (Y/b)^2 + (Z/c)^2$, has
the projected surface mass density given by the
following equation:
\begin{equation}
\kappa(\zeta) = \kappa_0 (1 + \zeta^2)^{(1-\nu)/2} \exp(-\zeta^2/R_t^2)
\Psi\left(\frac{1}{2},\frac{3}{2}-\frac{\nu}{2};\frac{1+\zeta^2}{R_t^2}\right),
\end{equation}
where $\Psi(a,b;z)$ is a confluent hypergeometric function and $\zeta$ is
given by equation~(3). Equation~(38) can be expanded in terms of `sple' 
models, and thus the $I$-functions of the model can be calculated in a
straightforward manner. 

Before we conclude this paper, below we discuss astrophysical motivations
and aspects of using the `cusp' double power-law model family as lens models,
which have not been emphasized in previous work on gravitational lensing.
For the `cusp' model given by equation~(1), the break radius and
the outer power-law index have been used as defining outer envelopes 
of systems of stellar objects (e.g.\ Jaffe 1983, Hernquist 1990, Dehnen 1993, 
Tremaine et al. 1994, Zhao 1996, Mu\~{n}oz et al.\ 2001), 
$\nu_o = 4$ being a preferred choice 
for the outer power-law index. Thus, the break 
radius would be a scale size of the system and the inner power-law
index would represent an overall radial profile within the scale size.
However, we also would like to use the break radius 
as a means to allow the radial profile of 
the lens to vary from a shallower inner profile, which is on average
 shallower than isothermal, to a steeper outer profile which is on average
similar to isothermal and not steeper than $\nu_o = 3$ in general. This would
be supported both by observational evidence and theoretical arguments.
First, rotation curves of spiral galaxies, dwarf galaxies, 
and low surface brightness galaxies generally rise in the inner
regions of the galaxies and are nearly flat beyond a turnover
radius (e.g.\ Kravtsov et al.\ 1998; Rubin, Waterman, \& Kenney 1999;
Sofue et al.\ 1999; van den Bosch et al.\ 2000; Swaters, Madore, \& Trewhella 
2000; de Blok et al.\ 2001). Rotation curves of small samples of 
ellipticals/S0 galaxies and blue compact galaxies show similar radial 
behaviors (Bertola \& Capaccioli 1975, 1977, 1978, Rubin, Peterson, \& Ford 
1980;  \"{O}stlin et al.\ 1999). These rotation curves imply 
shallower-than-isothermal inner profiles and steeper outer profiles which are 
similar to isothermal but generally not steeper than $\nu_o =3$. Second, 
{\it Hubble Space Telescope} luminosity profiles of the centers of early-type
galaxies show shallower-than-isothermal profiles (e.g.\ Rest et al.\ 2001; 
Faber et al.\ 1997; Gebhardt et al.\ 1996) virtually for all the galaxies 
imaged. These observations imply that, at least at the very centers of the 
galaxies where dark matter contribution to the total mass density is minimal, 
mass profiles are shallower than isothermal. Since we do not expect such 
shallow profiles to extend to large radii, varying radial profiles with 
relatively small break radii appear to be inevitable. 
Third, numerical simulations of halo formation in 
cold dark matter models predict mass distributions with varying 
radial profiles (e.g.\ Navarro, Frenk, White 1997; Moore et al.\ 1998; 
Jing \& Suto 2000; Ghigna et al.\ 2000) with  shallower-than-isothermal
inner profiles and an outer profile of $\nu_o =3$. Moreover, mass
distributions in the inner regions of the galaxies are expected to be 
modified due to the effects of the baryons (e.g.\ Blumenthal et al.\ 1986;
Mo, Mao, \& White 1998; Cole et al.\ 2000; Gonzalez et al.\ 2000; Kochanek \&
White 2001). It is thus of considerable interest to use gravitational lenses
to determine the inner profiles, break radii, and outer profiles 
(and eventually truncation radii, i.e., halo sizes) for
lensing (mostly early-type) galaxies. For this use of 
the outer power-law radial index, equation~(1) will, in general, have a 
diverging total mass without a well-defined extent of the galaxy, although 
it would not be an issue in applications of the
model to gravitational strong lensing, in which the extent of the dark matter
halo is assumed to be much larger than the critical radius of the lens. 
In order for the `cusp' model [eq.\ (1)] with $\nu_o \leq 3$ to have a 
well-defined extent and converging total mass, we could perhaps introduce
an envelope declining steeper than $\nu_o = 3$. In practice, it is likely 
that such a model is difficult to deal with mathematically. However, 
we could mimic an envelope using a subtraction term; in other words,
we could use the following modified version of equation~(1):
\begin{equation}
\rho(R) = \rho_0 R^{-\nu_i}[(1 + R^2)^{-(\nu_o-\nu_i)/2}
 - (R_t^2 + R^2)^{-(\nu_o-\nu_i)/2}].
\end{equation}
For large radius, the above model [eq.\ (39)] can be approximated by
$\rho(R) \propto R^{-(\nu_o+2)}$, which permits a finite total mass for any
$\nu_o > 1$. 

Finally, it is worth noting that in this paper we limited ourselves to
the families of two power-law mass distributions given by equations (1)
and (39) mainly because of straightforward mathematical tractabilities
of the models. The following model [see Hernquist (1990) and Zhao (1996)],
which is the generalized version of equation (39), 
would eventually be of use in future gravitational lens studies:
\begin{equation}
\rho(R) = \rho_0 R^{-\nu_i}[(1 + R^{\gamma})^{-(\nu_o-\nu_i)/\gamma}
 - (R_t^{\gamma} + R^{\gamma})^{-(\nu_o-\nu_i)/\gamma}],
\end{equation}
where parameter $\gamma$ represents the ``sharpness'' of the break.
However, until practical solutions to lensing by the model given by 
equation~(40) are obtained and observational data sensitive enough to 
constrain parameter $\gamma$ necessitate use of the model,
 equation (39) corresponding to $\gamma = 2$ in equation (40)
will be useful to study galactic radial profiles by applying the
model to gravitational lenses with the help of the calculational
method presented in this paper.

\acknowledgements
The author thanks Ian Browne, Chuck Keeton, Shude Mao, and Peter Wilkinson 
for encouragements, interests and useful comments on the manuscript. 
Chuck Keeton is also thanked for testing the code based on the results
of this paper, which helped to find a typographic error in the code. 
The author thanks David Turnshek for the past and on-going encouragements 
and support. He also would like to thank Wyn Evans and Neal Jackson for 
interests and comments on this work. Financial support for this work comes
from the universities of Manchester and Pittsburgh.

\appendix
\section{Calculation of $I$-functions of `Nuker' Model}
The $I$-functions of any surface mass density which is a function of
$\zeta$ [eq.\ (3)] are defined as follows:
\begin{eqnarray}
I^{(0)}(r) & = & \frac{1}{r \pi} \int_{0}^{2\pi} d\phi' 
 \int_0^r d r' r' \kappa(\zeta), \\
I^{(0')}(r) & = & \frac{r}{\pi} \int_{0}^{2\pi} d\phi' \kappa(\zeta), \\
I_{2m}^{(1)}(r) & = & \frac{r^{-1-2m}}{\pi \cos 2m\phi_0} B_{2m}(r)
 = \frac{r^{-1-2m}}{\pi \cos 2m\phi_0} \int_{0}^{2\pi} d\phi' \cos 2m\phi'
 \int_0^r d r' r'^{2m+1} \kappa(\zeta), \\
I_{2m}^{(2)}(r) & = & \frac{r^{-1+2m}}{\pi \cos 2m\phi_0} D_{2m}(r)
 = \frac{r^{-1+2m}}{\pi \cos 2m\phi_0} \int_{0}^{2\pi} d\phi' \cos 2m\phi'
 \int_r^{\infty} d r' r'^{-2m+1} \kappa(\zeta), \\
I_{2m}^{(3)}(r) & = & \frac{r}{2 \pi \cos 2m\phi_0}
 \int_{0}^{2\pi} d\phi' \cos 2m\phi' \kappa(\zeta), 
\end{eqnarray}
where $\zeta = \zeta(r',\phi') = (r'/r_b)[1 + e \cos 2(\phi' -\phi_0)]^{1/2}$
and $m =1,2,3,\ldots$\footnote{Notice that $d[I^{(0)}(r)]/dr =
[-I^{(0)}(r) + I^{(0')}(r)]/r$, and similarly for $I_{2m}^{(1)}(r)$ and
$I_{2m}^{(2)}(r)$.}

Now for the `Nuker' model [eq.\ (5)] for non-integer values of $\lambda$ we 
have 
\begin{eqnarray} 
B_{2m}(r) & = & \int_{0}^{2\pi} d\phi' \cos 2m\phi' \int_0^r d r' r'^{2m+1}
 \kappa^{\mbox{\scriptsize Nuker}}(r',\phi') \nonumber \\
   & = & \int_{0}^{2\pi} d\phi' \cos 2m\phi' \int_0^r d r' r'^{2m+1}
  \frac{[r'^2 g(\phi')]^{\lambda}}{[1 + r'^2 g(\phi')]^{\mu+1}} \nonumber \\
  & = & \frac{r^{2(m+1)}}{2(\lambda +m+1)} \int_{0}^{2\pi} d\phi' \cos 2m\phi'
  [g(\phi')r^2]^{\lambda} \nonumber \\
  &  & \times F[\mu+1,\lambda+m+1;\lambda+m+2; -g(\phi')r^2]  \nonumber \\
  & = & \frac{r^{2(\lambda +m+1)}}{2(\lambda +m+1)} \sum_{n=0}^{\infty}
  \frac{(-1)^n (\mu+1)_n (\lambda +m+1)_n}{(\lambda +m+2)_n n!} \nonumber \\
  &  & \times r^{2n}\int_{0}^{2\pi} d\phi'\cos 2m\phi' [g(\phi')]^{\lambda+n}
  \end{eqnarray}
where $g(\phi') \equiv P + Q \cos 2(\phi'-\phi_0) \equiv r_b^{-2} 
[1 + e \cos 2(\phi' - \phi_0)]$ ($e, Q \ge 0$) as defined in CKT.
The integral in the last line of equation (A6) can be evaluated,  using 
equation (B3) of CKT, as follows
\begin{equation}
\int_0^{2\pi} d\phi' \cos 2m\phi' [g(\phi')]^{\lambda+n}
  = 2\pi (-1)^m \cos(2m\phi_0) \frac{(P^2 -Q^2)^{(\lambda+n)/2}}
 {(1+\lambda+n)_m} P_{\lambda+n}^m \left( \frac{P}{\sqrt{P^2-Q^2}} \right).
\end{equation}
Now using equation (A4) of CKT and noting the definition of 
$I_{2m}^{(1)}(r)$ above [eq.\ (A3)], equation (17) for $i=1$ is obtained.

Similarly we have
\begin{eqnarray} 
D_{2m}(r) & = & \int_{0}^{2\pi} d\phi' \cos 2m\phi' \int_r^{\infty} 
d r' r'^{-2m+1}
  \frac{[r'^2 g(\phi')]^{\lambda}}{[1 + r'^2 g(\phi')]^{\mu+1}} \nonumber \\
 & = & \frac{r^{-2(m-\lambda+\mu)}}{2(m-\lambda+\mu)} 
\int_{0}^{2\pi} d\phi' \cos 2m\phi' [g(\phi')]^{\lambda-\mu-1}  \nonumber \\
 &  & \times F\left[\mu+1,m-\lambda+\mu;m-\lambda+\mu+1;-\frac{1}{g(\phi')r^2}
   \right].   
 \end{eqnarray}
The hypergeometic function in equation (A8) can be re-written, via a linear
transformation of hypergeometric functions (e.g.\ Gradshteyn \& Ryzhik 1994),
as follows
\[ F\left[\mu+1,m-\lambda+\mu;m-\lambda+\mu+1;-\frac{1}{g(\phi')r^2}\right]=\]
\begin{eqnarray}
 &  & \frac{\Gamma(m-\lambda+\mu+1)\Gamma(m-\lambda-1)}
    {\Gamma(m-\lambda+\mu) \Gamma(m-\lambda)} [g(\phi')r^2]^{\mu+1} 
    \nonumber \\
  &  &  \times F[\mu+1,-m+\lambda+1;-m+\lambda+2;-g(\phi')r^2] \nonumber \\
  &  & + \frac{\Gamma(m-\lambda+\mu+1)\Gamma(-m+\lambda+1)}{\Gamma(\mu+1)}
   [g(\phi')r^2]^{m-\lambda+\mu}
  \end{eqnarray} 
Substituting equation (A9) into equation (A8) and integrating over $\phi'$,
we find that the second term in equation (A9) vanishes. The rest of the 
calculation is very similar to that for $B_{2m}(r)$ and we obtain
equation (17) for $i=2$.
The function $I^{\mbox{\scriptsize Nuker}(3)}_{2m}(r;\lambda,\mu)$ [eq.\ (17)
for $i=3$] can also be obtained in a very similar way from the definition
[eq.\ (A5)].

\setcounter{equation}{0}
\section{Revisiting `SPLE' Model: Some New Expressions for $I$-Functions}
As described in \S 3, $I$-functions of the `cusp' model [eqs.\ (2a,b)]
are calculated using those of the `sple' model [eq.\ (4)] as well as those of
the `Nuker' model [eq.\ (5)] since the `cusp' model can be expanded in 
terms of the `sple' and `Nuker' models. Here we consider some new expressions 
for the $I$-functions of the `sple' model, which are useful for calculating 
$I$-functions of the `cusp' model.

The $I$-functions of the `sple' model obtained in CKT converge very quickly 
either for large or small radii compared with the core radius, which is
denoted by $r_b$ in equation (4). However, neither expression for the
$I^{(1)}$-function given in CKT converges sufficiently fast to be useful
for calculating the $I^{(1)}$-function of the `cusp' model for
$r \sim r_b$, although the $I^{(2)}$-function given in CKT is still useful.

We now obtain new expressions for the $I^{(1)}$- and $I^{(2)}$-functions of
the `sple' model which converge very quickly for $r \sim r_b$.
This is done via the following expansion of the `sple' model,
\begin{eqnarray}
\kappa^{\mbox{\scriptsize sple}} (r,\phi) & = &
\left\{ 1 + \left(\frac{r}{r_b}\right)^2 [1 + e \cos 2(\phi-\phi_0)]
\right\}^{-(\mu+1)} \nonumber \\
 & = & \left[ 1 + \left(\frac{r}{r_b}\right)^2 \right]^{-(\mu+1)}
 \left\{ 1 + \frac{(\frac{r}{r_b})^2}{1+(\frac{r}{r_b})^2} 
 e \cos 2(\phi-\phi_0) \right\}^{-(\mu+1)}  \nonumber \\
 & = & \sum_{k=0}^{\infty} \frac{(-1)^k (\mu+1)_k}{k!} 
 \frac{(\frac{r}{r_b})^{2k}}{[1+(\frac{r}{r_b})^2]^{k+\mu+1}} 
 e^k [\cos 2(\phi-\phi_0)]^k.
\end{eqnarray}
After straightforward algebra using equation (B1), we obtain
\begin{eqnarray}
I_{2m}^{{\mbox{\scriptsize sple}}(1)}(r;\mu) & = & \frac{r}{\pi} 
 (-te)^m (1-t)^{\mu+1} 
 \sum_{k=0}^{\infty} \frac{\Gamma(k+m+\mu+1)}{\Gamma(\mu+1)\Gamma(k+m+1)}
 \frac{1+(-1)^k}{k+2m+1} \nonumber \\
   &  & \times (te)^k {\mbox{Icos}}(k+m,m)
   F(k+m+\mu+1,1;k+2m+2;t), \\
I_{2m}^{{\mbox{\scriptsize sple}}(2)}(r;\mu) & = & \frac{r}{\pi}
 (-te)^m (1-t)^{\mu+1} \frac{1}{(\mu +m)t}
  \sum_{k=0}^{\infty} \frac{\Gamma(k+m+\mu+1)}{\Gamma(\mu+1)\Gamma(k+m+1)}
  \nonumber \\
 &  & \times [1+(-1)^k] e^k {\mbox{Icos}}(k+m,m)
   F(-k,m+\mu;m+\mu+1;1-t),  
\end{eqnarray}
where $t$ and Icos$(k,m)$ are given in equation (30a). 

\setcounter{equation}{0}
\section{Prescriptions for Evaluating Non-trivial $I$-Functions}
All the $I$-functions obtained in CKT and in this paper are well-defined
converging series. Numerical implementation of them is relatively 
straightforward. However, a few of them need to be dealt with carefully due
to their non-trivially converging properties for part of the parameter space. 
Below we discuss these non-trivial $I$-functions and give some mathematical
prescriptions which are useful for numerically evaluating them.

Let us first consider the $I^{(2)}$-function of the `sple' model obtained in 
CKT, which is given by
\begin{eqnarray}
I_{2m}^{{\mbox{\scriptsize sple}}(2)}(r;\mu) & = & h(r) 
[-\sqrt{\varepsilon_1(r)}]^m [\varepsilon_2(r)]^{\mu} 
\frac{\Gamma(m+\mu)}{\Gamma(m+1) \Gamma(\mu+1)} \nonumber \\
 &  & \times \sum_{k=0}^{\infty} [\varepsilon_2(r)]^k 
  F[m-k-\mu,-k-\mu;m+1;\varepsilon_1(r)].
\end{eqnarray}
While this function converges quickly for $r \gtrsim r_b$ since 
$\varepsilon_2(r) \rightarrow 0$ for $r \gtrsim r_b$, it converges slowly
for $r < r_b$ especially as $r/r_b \rightarrow 0$. 
The latter case (i.e.\ $r < r_b$) is irrelevant in most applications of the
 `sple' model since observed images are normally found well outside any lens 
core radii. However, as can be seen in equations (14) and (23), evaluation of
the $I^{(2)}$-function of the `cusp' model for $r < r_b$ requires evaluation
of the $I^{(2)}$-function of the `sple' model [eq.\ (C1)] for $r < r_b$ 
multiple times. It is thus important to evaluate equation
(C1) fast and accurately for $r < r_b$. 
Fast and accurate evaluation of equation (C1) can be done by expanding the
hypergeomtric function and noting that infinite sum for each term in 
the hypergeometric function can be broken down to terms including 
\begin{equation}
S_n(x) = \sum_{k=1}^{\infty} k^n x^k \hspace{1em} (n=0,1,2,\ldots),
\end{equation}
which can be analytically evaluated using the following recursion relation
\begin{equation}
S_n(x) =\frac{-1}{1-x} \sum_{k=0}^{n-1} (-1)^{n-k} 
\left(\begin{array}{c} n \\ k \end{array}\right) S_k(x)
\end{equation}
and $S_0 (x) = x/(1-x)$.

As noted in \S 3, the $I^{(1)}$-function given by equation (30a) of the
`cusp' model requires a special treatment. Equation (30a), which was 
obtained by  substituting equation (B2) into equation (26), has a slow 
convergence over $l$ although its convergence over $k$ is fast. 
This slow convergence problem
cannot be overcome using equation (27). In fact, the $I^{(1)}$-function given
by equation (30a), which is needed for $r \sim r_b$, is the only true slowly
converging series out of all the $I$-functions given in CKT and in 
this paper. Let the $l$-th term in the infinite sum over $l$ in equation (30a)
be denoted by $u_l$, i.e.,
\begin{equation}
u_l = D_l(\nu_i,\nu_o) (1-t)^l
    \frac{\Gamma(l+k+m+\mu+1)}{\Gamma(l+\mu+1)}
      F(l+k+m+\mu+1,1;k+2m+2;t).
\end{equation}
As $l \rightarrow \infty$, we have $u_l/u_{l+1} \rightarrow 1$ and
\begin{equation}
l \left( \frac{u_l}{u_{l+1}} -1 \right) \rightarrow a > 1,
\end{equation}
which implies  
\begin{equation}
 u_l \rightarrow \frac{l-1}{a+l-1} u_{l-1}.
\end{equation}
Equation (C6) can then be used to approximate the summation after $l_0 \gg 1$
as follows,
\begin{equation}
\sum_{l=l_0}^{\infty} u_l\rightarrow u_1 \frac{a}{a-1}-\sum_{l=1}^{l_0-1}u_l.
\end{equation}

\newpage
\centerline{\bf Table 1}
\centerline{\bf Summary of Mass Models}
\tabskip=1.2em
\halign to 
\hsize{\hfil#\hfil&\hfil#\hfil&\hfil#\hfil&\hfil#\hfil
\cr
\noalign{\vskip6pt\hrule\vskip3pt\hrule\vskip6pt}
name  &  meaning  &  projected density & 3-dimensional density \cr
\noalign{\vskip6pt\hrule\vskip6pt}
SPLE & softened power-law ellipsoid &
 Equation (4)  &  $\rho(R)= \rho_0 (1+R^2)^{-\nu/2}$ $^1$  \cr
Nuker & ``Nuker-law'' model  & Equation (5) &  Not Applicable \cr
Cusp & cusped two power-law model & Equations (2a,b)
 &  Equation (1) \cr
\noalign{\vskip3pt\hrule}
}

$^1$ See equation (1) for the definition of $R$.
\end{document}